\documentclass[11pt]{amsart}
\usepackage{geometry}                
\usepackage{graphicx}
\usepackage{amssymb}
\usepackage{epstopdf}
\title{An empirical partition function for two-dimensional Ising Model in an external magnetic field}
\author{Rong Qiang Wei}
\address{College of Earth and Planetary Sciences, University of Chinese Academy of Sciences, Beijing, PRC, 100049}
\email{wrq1973@ucas.edu.cn}
\date{}

\begin{document}
\maketitle

\begin{abstract}
     There is no an accepted exact partition function (PF) for the two-dimensional (2D) Ising model with a non-zero external magnetic field to our knowledge. Here we infer an empirical PF for such an Ising model. We compare the PFs for two finite-size Ising lattices ($4\times 4$ and $4\times 6$) from this empirical PF with those from Wei (2018) (Wei, R.Q., 2018. An exact solution to the partition function of the finite-size Ising Model, arXiv: General Physics: 1805.01366.), and find that they are consistent very well. Based on this empirical PF, we further analyze and calculate the thermodynamic functions (heat capacity, magnetization, susceptibility) of this 2D Ising model and discuss the model's singularity semiquantitatively. Analysis and calculations from this PF show that they are coincident with those from other related studies; Especially the 2D Ising model in an external magnetic field has spontaneous magnetization (SM) calling the phenomenon at the critical temperature a phase transition, and the SM decreases with the increasing temperature. However, the decreasing variation of the SM here is different from that obtained by Yang (1952) from the 2D Ising model in a weak magnetic field.  
  
\end{abstract}

{\hspace{2.2em}\small Keywords:}

{\hspace{2.2em}\tiny 2D Ising model, External magnetic field, Empirical partition function, Spontaneous magnetization}

\section{Introduction}

For the 2D Ising model, most studies focus on those in the absence of a non-zero external magnetic field. Among them, the famous exact PF ($O_{_{2d}}(0,T)$) was obtained by Onsager (1944) using the transfer-matrix method, 

\begin{equation}\label{Onsager}
\log O_{_{2d}}(0,T)=\log(2\cosh 2z)+\frac{1}{2\pi}\int_0^\pi {\rm{d}}\phi\log\frac{1}{2}\left(1+\sqrt{1-\kappa^2\sin^2\phi}\right)
\end{equation}
where $z=\beta\epsilon$, $\beta=\frac{1}{kT}$, $k$ is Boltzmann constant, $T$ the temperature, $\epsilon$ the interaction energy (and we assume that $\epsilon>0$ here), and $$\kappa=\frac{2}{\cosh 2z \coth 2z}$$

Other 2D lattices, eg., the triangular, honeycomb, and decorated lattices have all been considered. For example, Houtappel (1950) evaluate the exact PF on the triangular ($Q_{\rm{triangle}}$) and honeycomb ($Q_{\rm{honey}}$) lattices by a method quite closely related to transfer-matrix, 

\begin{equation}\label{Houtappel}
\begin{array}{lr}
\log Q_{\rm{honey}} &={\frac{1}{{16{\pi ^2}}}\int_0^{2\pi } {\int_0^{2\pi } {\log \frac{1}{2}\{ {{\cosh }^3}2z + 1 - {{\sinh }^2}2z[\cos {\omega _1}} } }\\
{}&{ + \cos {\omega _2} + \cos ({\omega _1} + {\omega _2})]\} {\rm{d}}{\omega _1}{\rm{d}}{\omega _2}}
\end{array}
\end{equation}

\begin{equation}\label{Houtappel2}
\begin{array}{lr}
\log Q_{\rm{triangle}}&={\frac{1}{{8{\pi ^2}}}\int_0^{2\pi } {\int_0^{2\pi } {\log \{ {{\cosh }^3}2z + {{\sinh }^3}2z - \sinh 2z[\cos {\omega _1}} } }\\
{}&{ + \cos {\omega _2} + \cos ({\omega _1} + {\omega _2})]\} {\rm{d}}{\omega _1}{\rm{d}}{\omega _2}}
\end{array}
\end{equation}

However,  there is still no an accepted exact solution for the 2D Ising model with a non-zero external magnetic field, especially the one which is both elegant and exact like Eq. (\ref{Onsager})- Eq. (\ref{Houtappel2}).  By using the Hubbard-Stratonovich transformation, Wei (2018) presented a simple but exact solution to the PF for such a finite-size 2D Ising model,

\begin{equation}\label{wei2d}
\begin{array}{ll}
Q(H,T)=&{\exp \left( {\frac{1}{2}{{\bf{K}}_0}{{\bf{K}}^{ - 1}}{\bf{K}}_0^T + N\beta h} \right)}\\
{}&{ + \sum\limits_{\alpha  = 1}^N {\exp \left[ {\frac{1}{2}({{\bf{K}}_0} - 2{{\bf{K}}_\alpha }){{\bf{K}}^{ - 1}}{{({{\bf{K}}_0} - 2{{\bf{K}}_\alpha })}^T} + (N - 2)\beta h} \right]} }\\
{}&{ + \sum\limits_{\alpha ,\beta  = 1(\alpha  < \beta )}^N {\exp \left[ {\frac{1}{2}({{\bf{K}}_0} - 2{{\bf{K}}_\alpha } - 2{{\bf{K}}_\beta }){{\bf{K}}^{ - 1}}{{({{\bf{K}}_0} - 2{{\bf{K}}_\alpha } - 2{{\bf{K}}_\beta })}^T} + (N - 4)\beta h} \right]} }\\
{}& + \sum\limits_{\alpha ,\beta ,\gamma  = 1(\alpha  < \beta  < \gamma )}^N \exp \left[ {\frac{1}{2}({{\bf{K}}_0} - 2{{\bf{K}}_\alpha } - 2{{\bf{K}}_\beta } - 2{{\bf{K}}_\gamma }){{\bf{K}}^{ - 1}}{{({{\bf{K}}_0} - 2{{\bf{K}}_\alpha } - 2{{\bf{K}}_\beta } - 2{{\bf{K}}_\gamma })}^T}}\right. \\
{}&  \hspace{30em} +(N - 6)\beta h] \\
{}&{+......}\\
{}&{+\exp \left[ {\frac{1}{2}( - {{\bf{K}}_0}){{\bf{K}}^{ - 1}}( - {\bf{K}}_0^T) - N\beta h} \right]}
\end{array}
\end{equation}
where $h=\mu H$ is the Zeeman energy associated with an external magnetic field $H$, $\mu$ the permeability. $\bf{K}$ denotes the exchange interaction matrix, ${\bf{K}}_0=\sum\limits_i^N {\bf K}_i$,  and {\bf K} is (Dixon et al., 2001),

\[{\bf K} = \left[ {\begin{array}{*{20}{c}}
{\bf A}&{\bf I}&0&0& \cdot & \cdot & \cdot &{\bf I}\\
{\bf I}&{\bf A}&{\bf I}&0& \cdot & \cdot & \cdot &0\\
0&{\bf I}&{\bf A}&{\bf I}& \cdot & \cdot & \cdot &{}\\
0&0&{\bf I}&{}&{}&{}&{}&{\bf I}\\
{\bf I}&0& \cdot & \cdot & \cdot &{}&{\bf I}&{\bf A}
\end{array}} \right]\]
where {\bf A} is, 

\[{\bf A} =z\times \left[ {\begin{array}{*{20}{c}}
0&1&0&0& \cdot & \cdot & \cdot &1\\
1&0&1&0&{}&{}&{}&{}\\
0&1&0&1&{}&{}&{}&{}\\
0&0&1&0&{}&{}&{}&\cdot\\
 \cdot &\cdot&\cdot&{}&{}&{}&{}&\cdot\\
 \cdot &\cdot&\cdot&{}&{}&{}&{}&\cdot\\
 \cdot &\cdot&\cdot&{}&{}&{}&{}&1\\
1&0&0&{}&{}&{}&1&0
\end{array}} \right]\]

However, Eq. (\ref{wei2d})  looks formidable and unintuitive because it is expressed in a sum of $2^N$ exponential functions. It is only practicable to calculate the PF for the 2D Ising model with a small $N$. 

In the absence of a practicable exact solution, an empirical PF, which is under some reliable constraints, is one of the possible ways to understand the 2D Ising model with a non-zero external magnetic field. Here we infer such an empirical PF, and investigate its thermodynamic functions and discuss their singularity semiquantitatively. 

\section{An empirical PF for the 2D Ising Model}\label{sec2}

\ \ \ 

We start with rewriting the PF for the 1D Ising model in an external magnetic field $h$. For this model, the PF is (Kramers and Wannier, 1941),

\begin{equation}\label{KW1-orig}
\frac{1}{N}\log Q_{_{1d}}(h,T)=\log \left[\exp (z)\cosh \beta h + \sqrt{\exp(2z)\sinh^2 \beta h+\exp(-2z)}\right]
\end{equation}

For $h=0$, 
 \begin{equation}\label{KW1-orig2}
\begin{array}{ll}
 \frac{1}{N}\log Q_{_{1d}}(0,T) &={\log (2\cosh z)}\\
{}&{}\\
  &={\frac{1}{2}\log (2\sinh z) + \frac{1}{2}\ln (2\coth z\cosh z)}\\
{}&{}\\
  &\approx{\frac{1}{2}\log (2\sinh z) + \frac{1}{2}{{\cosh }^{ - 1}}(\coth z\cosh z)}\\
{}&{}\\
  &={\frac{1}{2}\log (2\sinh z) + \frac{1}{{2\pi }}\int_0^\pi  {\log (2\coth z\cosh z - 2\cos {\omega _1}){\rm{d}}{\omega _1}}}
\end{array}
\end{equation}
where $\cosh^{-1}z\approx \log 2z \ \ (z \ge 1)$ and $\cosh^{-1}z=\frac{1}{\pi}\int_0^\pi \log(2z-2\cos\omega_1){\rm{d}}\omega_1$ are used (See details in Appendix A).

Similarly, 

\begin{equation}\label{KW1}
\begin{array}{ll}
\frac{1}{N}\log Q_{_{1d}}(h,T) &=\log \left[\exp (z)\cosh \beta h + \sqrt{\exp(2z)\sinh^2 \beta h+\exp(-2z)}\right]\\
{} &=\frac{1}{2}\log \left[\exp (z)\cosh \beta h + \sqrt{\exp(2z)\sinh^2 \beta h+\exp(-2z)}\right]^2\\
{}&=\frac{1}{2}\log(2\sinh z)+\frac{1}{2}\log \frac{\left[\exp (z)\cosh \beta h + \sqrt{\exp(2z)\sinh^2 \beta h+\exp(-2z)}\right]^2}{2\sinh z}\\
  {}&{}\\
  &\approx{\frac{1}{2}\log (2\sinh z) + \frac{1}{2}{{\cosh }^{ - 1}}\left\{ \frac{\left[\exp (z)\cosh \beta h + \sqrt{\exp(2z)\sinh^2 \beta h+\exp(-2z)}\right]^2}{4\sinh z}\right\}}\\
{}&{}\\
  &=\frac{1}{2}\log (2\sinh z) + \frac{1}{{2\pi }}\int_0^\pi  \log (2{\frac{\left[\exp (z)\cosh \beta h + \sqrt{\exp(2z)\sinh^2 \beta h+\exp(-2z)}\right]^2}{4\sinh z}} \\
  {}&\hspace{24em} - 2\cos {\omega _1}){\rm{d}}{\omega _1}
\end{array}
\end{equation}

\ \ \ 

According to Huang (1987) and Martin (1991), the PF for 2D Ising model in the absence of an external field is,

 \begin{equation}\label{Onsager1}
\begin{array}{ll}
\frac{\log Q_{_{2d}}(0,T)}{M \times N}&{ = \frac{1}{2}\log (2\sinh 2z) + \frac{1}{{2{\pi ^2}}}\int_0^\pi  {\int_0^\pi  {\log (2\cosh 2z\coth 2z - 2\cos {\omega _1} - 2\cos {\omega _2}){\rm{d}}{\omega _1}{\rm{d}}{\omega _2}} } }\\
\end{array}
\end{equation}

Analogous to Eq. (\ref{KW1-orig})-Eq. (\ref{KW1}), we infer the PF for 2D Ising model with a non-zero external field is,

 \begin{equation}\label{wei2d2}
\begin{array}{lr}
\frac{\log Q_{_{2d}}(h,T)}{M \times N}& = \frac{1}{2}\log (2\sinh 2z) + \frac{1}{{2{\pi ^2}}}\int_0^\pi  \int_0^\pi \log \{2{\frac{\left[\exp (2z)\cosh \beta h + \sqrt{\exp(4z)\sinh^2 \beta h+\exp(-4z)}\right]^2}{4\sinh 2z}}\\
{}& - 2\cos {\omega _1} - 2\cos {\omega _2}\}{\rm{d}}{\omega _1}{\rm{d}}{\omega _2} \\
\end{array}
\end{equation}

To test Eq. (\ref{KW1}) and (\ref{wei2d2}), we discretise the corresponding integral to a sum,

\begin{equation}\label{KW1-sum}
\begin{array}{lr}
\frac{1}{N}\log {Q_{1d}}(h,T) =& \log \left[ \exp (z)\cosh \beta h + \sqrt{\exp(2z)\sinh^2 \beta h+\exp(-2z)} \right]\\
& + \frac{1}{{N}}\sum\limits_r^N {\log [1 - J_1\cos \frac{{2\pi r}}{N}]}
\end{array}
\end{equation}
where $J_1$ is,
\[ J_1=\frac{4\sinh z}{\left[\exp (z)\cosh \beta h + \sqrt{\exp(2z)\sinh^2 \beta h+\exp(-2z)}\right]^2}\]

\begin{equation}\label{wei2d2-sum}
\begin{array}{lr}
\frac{1}{M\times N}\log {Q_{2d}}(h,T) =& \log \left[\exp (2z)\cosh \beta h + \sqrt{\exp(4z)\sinh^2 \beta h+\exp(-4z)} \right]\\
& + \frac{1}{{MN}}\sum\limits_r^M\sum\limits_s^N {\log [1 - J_2(\cos \frac{{2\pi r}}{M}+\cos \frac{{2\pi s}}{N})]}
\end{array}
\end{equation}
where $J_2$ is,
\[ J_2=\frac{4\sinh 2z}{\left[\exp (2z)\cosh \beta h + \sqrt{\exp(4z)\sinh^2 \beta h+\exp(-4z)}\right]^2}\]

Of course, Eq. (\ref{KW1}) or (\ref{wei2d2}) can be discretised to a product, for example,

\begin{equation}\label{wei2d-prod}
\begin{array}{lr}
{Q_{2d}}(h,T) =& \left[\exp (2z)\cosh \beta h + \sqrt{\exp(4z)\sinh^2 \beta h+\exp(-4z)} \right]^{M\times N}\\
& \times \prod\limits_{r = 1}^M\prod\limits_{s = 1}^N [1 - J_2(\cos \frac{{2\pi r}}{M}+\cos \frac{{2\pi s}}{N})]
\end{array}
\end{equation}

\begin{figure}[htb]
 \centering
 \includegraphics[scale=0.5]{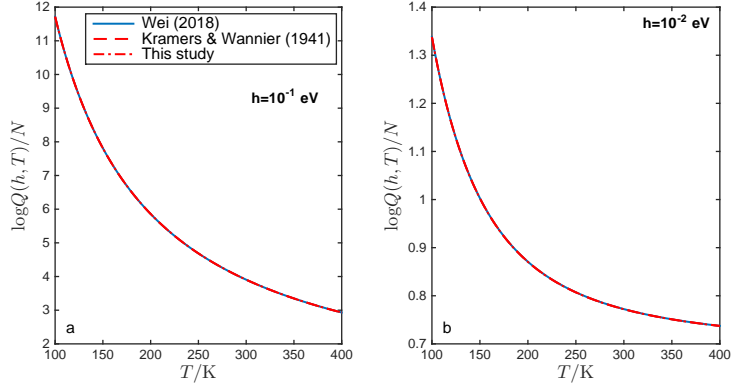}
 \caption{\footnotesize The partition function (PF) vs. temperature calculated from Eq. (\ref{wei2d}), Eq. (\ref{KW1-orig}), Eq. (\ref{KW1-sum}) for $\epsilon=1.0\times 10^{-3}\mbox{eV}$, respectively. a. N=10; b. N=24.}
\label{fig1}
\end{figure}

Fig. \ref{fig1} shows the PFs vs. temperature for the finite-size 1D Ising model calculated from Eq. (\ref{wei2d}), Eq. (\ref{KW1-orig}), Eq. (\ref{KW1-sum}). It can be seen that these three PFs are fully consistent, which shows that our rewriting the PF of Kramers and Wannier (1941), and the PF from Wei (2018) are correct. 

\begin{figure}[htb]
 \centering
 \includegraphics[scale=0.5]{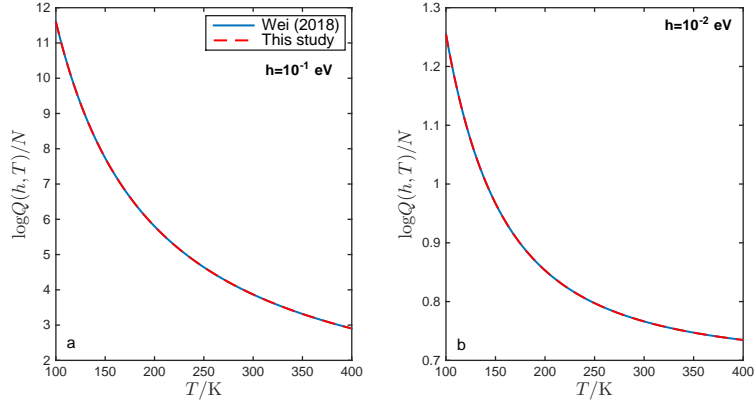}
 \caption{\footnotesize The PFs vs. temperature calculated from Eq. (\ref{wei2d}), Eq. (\ref{wei2d2-sum}) for $\epsilon=1.0\times 10^{-3}\mbox{eV}$, respectively.  a. $M\times N=4\times4$; b. $M\times N=4\times 6$.}
\label{fig2}
\end{figure}

Fig. \ref{fig2} shows the PFs vs. temperature for the finite-size 2D Ising model calculated from Eq. (\ref{wei2d}), Eq. (\ref{wei2d2-sum}). It can be seen that these two PFs are fully consistent, which show numerically that the PF (Eq. (\ref{wei2d2})) is correct. Because it is not proved rigorously but only supported by the results from two 2D lattices at present, we call it empirical PF for the 2D Ising model with a non-zero external magnetic field. It should be pointed out that there are other empirical PFs for this Ising model.

\section{Thermodynamic functions from the empirical PF}
\subsection{Specific heat and the critical temperature}
With Eq. (\ref{wei2d2}), we can discuss the thermodynamic functions of the 2D Ising model with a non-zero external magnetic field in two ways. One way is analogous to Mccoy and Wu (1973), or Huang (1987), in which we rewrite Eq. (\ref{wei2d2}) as,

\begin{equation}\label{wei2d2-another}
\begin{array}{lr}
{\frac{1}{N}\log Q(h,T)}&{ = \log \left[ {\exp (2z)\cosh \beta h + \sqrt {\exp (4z){{\sinh }^2}\beta h + \exp ( - 4z)} } \right]}\\
{}&{ + \frac{1}{{2\pi }}\int_0^\pi  {{\rm{d}}\phi \log \frac{1}{2}(1 + \sqrt {1 - {\kappa ^2}{{\sin }^2}\phi } )} }
\end{array}
\end{equation}
where $\kappa$ is, 
\[\kappa=\frac{8\sinh 2z}{\left[\exp (2z)\cosh \beta h + \sqrt{\exp(4z)\sinh^2 \beta h+\exp(-4z)}\right]^2}\]

Hence the Helmholtz free energy per spin $a_{_{I}}(h,T)$:

\begin{equation}\label{Helmholtz-free-energy}
\begin{array}{lr}
\beta a_{_{I}}(h,T)&{ = -\log \left[ {\exp (2z)\cosh \beta h + \sqrt {\exp (4z){{\sinh }^2}\beta h + \exp ( - 4z)} } \right]}\\
{}&{ - \frac{1}{{2\pi }}\int_0^\pi  {{\rm{d}}\phi \log \frac{1}{2}(1 + \sqrt {1 - {\kappa ^2}{{\sin }^2}\phi } )} }
\end{array}
\end{equation}

From Eq. (\ref{Helmholtz-free-energy}), one can obtain the internal energy, heat capacity, magnetization, susceptibility of the 2D Ising model with a non-zero external magnetic field.  The approach is similar to that in Mccoy and Wu (1973) or Huang (1987), but the results are more complicated. It should be pointed out that the internal energy is continuous at the critical temperature in this case.  

The other way is using lattice Green function (LGF) of the 2D lattice.  For convenience, we rewrite Eq. (\ref{wei2d2}) as,

 \begin{equation}\label{wei2dforLGF}
\begin{array}{ll}
\frac{\log Q_{_{2d}}(h,T)}{N}& \sim \frac{1}{{2{\pi ^2}}}\int_0^\pi  \int_0^\pi \log \{2{\frac{\left[\exp (2z)\cosh \beta h + \sqrt{\exp(4z)\sinh^2 \beta h+\exp(-4z)}\right]^2}{4\sinh 2z}}\\
{}&{}\\
{}& - 2\cos {\omega _1} - 2\cos {\omega _2}\}{\rm{d}}{\omega _1}{\rm{d}}{\omega _2} \\
{}&{}\\
{}& = \frac{1}{{2{\pi ^2}}}\int_0^\pi  \int_0^\pi \log (2f - 2\cos {\omega _1} - 2\cos {\omega _2}\}{\rm{d}}{\omega _1}{\rm{d}}{\omega _2}
\end{array}
\end{equation}
where "$\sim$" means the analytical $\frac{1}{2}\log (2\sinh 2z)$ is not taken into account, and:
\[f=\frac{\left[\exp (2z)\cosh \beta h + \sqrt{\exp(4z)\sinh^2 \beta h+\exp(-4z)}\right]^2}{4\sinh 2z}\]

Then the internal energy $U$ is,

\begin{equation}\label{internal-energy}
\begin{array}{ll}
 U&=-\frac{\partial }{\partial\beta}[\frac{1}{N}\log Q_{_{2d}}(h,T)] \\
{}&{}\\
{}&\sim  \frac{1}{2{\pi ^2}} \int_0^\pi \int_0^\pi \frac{\frac{\partial f}{\partial \beta}{\rm{d}}{\omega _1}{\rm{d}}{\omega _2}}{f -(\cos {\omega _1} + \cos {\omega _2})}\\
{}&{}\\
{}&=\frac{1}{2\pi^3 f} K(\frac{2}{f})\frac{\partial f}{\partial \beta}
\end{array}
\end{equation}
where $K(\cdot)$ the complete elliptic integral of the first kind. 

Eq. (\ref{internal-energy}) can be analyzed qualitatively and simply to avoid the discussing the complete elliptic integral of the first kind $K(\cdot)$.  For this purpose, we adopt the analytic continuation formula for this equation in the immediate neighbourhood of the point $2/f = 1$ by Joyce (2003), which reads,

\begin{equation}\label{internal-energy2}
\frac{1}{{\pi ^2}}\int_0^\pi  \int_0^\pi  \frac{{{\rm{d}}{\omega _1}{\rm{d}}{\omega _2}}}{f -{(\cos {\omega _1} + \cos {\omega _2})} }=\sum\limits_{n = 0}^\infty  {{C_n}{{(f - 2)}^n}}  + {\log(f-2)}\sum\limits_{n = 0}^\infty  {{D_n}} {(f - 2)^n}
\end{equation}
where $C_n$, $D_n$ are constants satisfy some recurrence relations.

If we let $C(f)=\sum\limits_{n = 0}^\infty  C_n(f - 2)^n$, $D(f)=\sum\limits_{n = 0}^\infty  D_n(f - 2)^n$, then

\begin{equation}\label{internal-energy2}
U\sim \frac{1}{{\pi ^2}}\int_0^\pi  \int_0^\pi \frac{f'{{\rm{d}}{\omega _1}{\rm{d}}{\omega _2}}}{f -{(\cos {\omega _1} + \cos {\omega _2})} }=f'C(f)\left[1+\log(f-2)\frac{D(f)}{C(f)}\right]
\end{equation}

Thus the specific heat $C(H,T)$ is,

\begin{equation}\label{specific-heat}
\begin{array}{ll}
C(H,T)=-k\beta^2\frac{\partial U}{\partial\beta}&\sim \left[f''C(f)+(f')^2C'(f)\right]\left[1+\log(f-2)\frac{D(f)}{C(f)}\right]\\
{}&{}\\
{}&+f'C(f)\left\{\frac{f'D(f)}{C(f)(f-2)}+\log(f-2)\left[\frac{D(f)}{C(f)}\right]'\right\}
\end{array}
\end{equation}
where "$'$" means the derivative to $\beta$.

As has been pointed out that the internal energy is continuous at critical temperature, $\log(f-2)\frac{D(f)}{C(f)}$ is continuous as $f\to 2$. As long as $\left[\frac{D(f)}{C(f)}\right]' \nrightarrow 0$ when $f\to 2$, the specific heat $C(H,T)$ approaches infinity logarithmically at the critical temperature, like that in the 2D Ising model with a zero external field. And the critical temperature $T_c$ is such that,

\[
f=\frac{\left[\exp (2z)\cosh \beta h + \sqrt{\exp(4z)\sinh^2 \beta h+\exp(-4z)}\right]^2}{4\sinh 2z}=2
\]

\subsection{Magnetization}

One can also obtain the magnetization $M(h,T)$ as, 

\begin{equation}\label{magnetization}
\begin{array}{ll}
 M(h,T)&=\frac{\partial }{\partial h}[\frac{1}{N}\log Q_{_{2d}}(h,T)] \\
{}&{}\\
{}&=  \frac{1}{2{\pi ^2}} \int_0^\pi \int_0^\pi \frac{\frac{\partial f}{\partial h}{\rm{d}}{\omega _1}{\rm{d}}{\omega _2}}{f -(\cos {\omega _1} + \cos {\omega _2})}\\
{}&{}\\
{}&=\frac{1}{2\pi^3 f} K(\frac{2}{f})\frac{\partial f}{\partial h}\\
{}&{}\\
{}&=\frac{\beta}{\pi^3}K(\frac{2}{f})\frac{\sinh \beta h}{\sqrt{\sinh^2\beta h+\exp(-8z)}}\\
\end{array}
\end{equation}

If $T\to 0$, $\exp(-8z)=\exp(-\frac{8\epsilon}{kT})\to 0$, 

\begin{equation}\label{magnetization1}
M(h,T) = \left\{ {\begin{array}{ll}
{\frac{\beta}{{{\pi ^3}}}K(\frac{2}{f})}&{h > 0}\\
{ - \frac{\beta}{{{\pi ^3}}}K(\frac{2}{f})}&{h < 0}
\end{array}} \right.
\end{equation}
where obviously $T<T_c$.

Eq. (\ref{magnetization1}) holds when $h\to 0$, and

\begin{equation}\label{magnetization2}
M(0,T) = \left\{ {\begin{array}{ll}
{\frac{\beta}{{{\pi ^3}}}K(\frac{2\sinh 2z}{\cosh^2 2z})}&{h \to 0^+}\\
{ - \frac{\beta}{{{\pi ^3}}}K(\frac{2\sinh 2z}{\cosh^2 2z})}&{h \to 0^-}
\end{array}} \right.
\end{equation}

If $\exp(-\frac{8\epsilon}{kT})\nrightarrow 0$, especially for $T>T_c$, $K$ is finite, and 

\begin{equation}\label{magnetization3}
{\begin{array}{ll}
M(h,T) \to 0&{h \to 0}
\end{array}}
\end{equation}

For the case of $T=T_c$, additional discussion is required because $K(2/f)\vert_{_{T=T_c}}$ has a singularity.

Eq. (\ref{magnetization2})-(\ref{magnetization3}) show that the 2D Ising model has spontaneous magnetization (SM).  This implies that a phase transition occurs at $T=T_c$, but the SM from the empirical PF here is different from Eq. (\ref{Yang1952}) derived by Yang (1952) from the 2D Ising model in a weak magnetic field. We will see this clearly from the corresponding Fig. \ref{fig6} in the next subsection.

\begin{equation}\label{Yang1952}
M(0,T) = \left\{ {\begin{array}{ll}
0&{T > {T_c}}\\
{{{\left[ {1 - {{\sinh }^{ - 4}}2z} \right]}^{\frac{1}{8}}}}&{T < {T_c}}
\end{array}} \right.
\end{equation}

Furthermore, one can discuss and analyze the susceptibility, but the corresponding result is complicated which is related to the derivative of the complete elliptic integral of the first kind. We can take a peek at it from the calculations (Fig. \ref{fig5}) below.

\subsection{Calculation of the thermodynamic functions}

\begin{figure}[htb]
 \centering
 \includegraphics[scale=0.5]{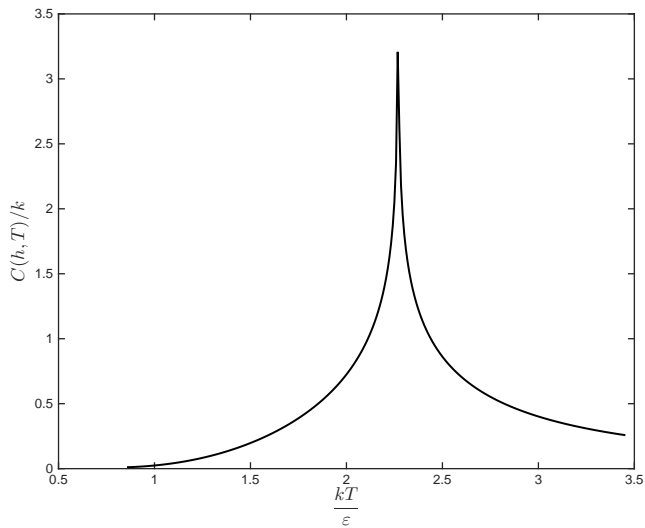}
 \caption{\footnotesize The heat capacity from the empirical PF. $\epsilon=1.0\times 10^{-2}\mbox{\ eV}$; $h=\mu H=1.0\times 10^{-5}\mbox{eV}$; $T$ is from 100 to 400 $\mbox{\ K}$.}
\label{fig3}
\end{figure}

\begin{figure}[htb]
 \centering
 \includegraphics[scale=0.5]{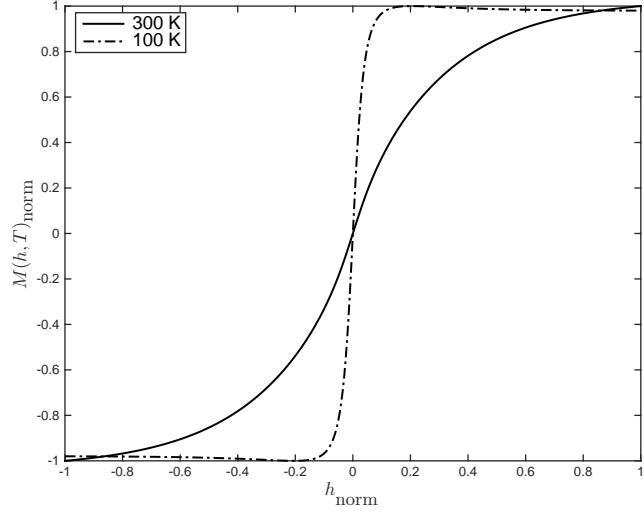}
 \caption{\footnotesize The magnetization from the empirical PF. $\epsilon=1.0\times 10^{-2}\mbox{eV}$; $\mu H$ is from $-0.02$ to $0.02\mbox{\ eV}$. The magnetization and the external magnetic field are normalized by their maximum, respectively. }
\label{fig4}
\end{figure}

\begin{figure}[htb]
 \centering
 \includegraphics[scale=0.5]{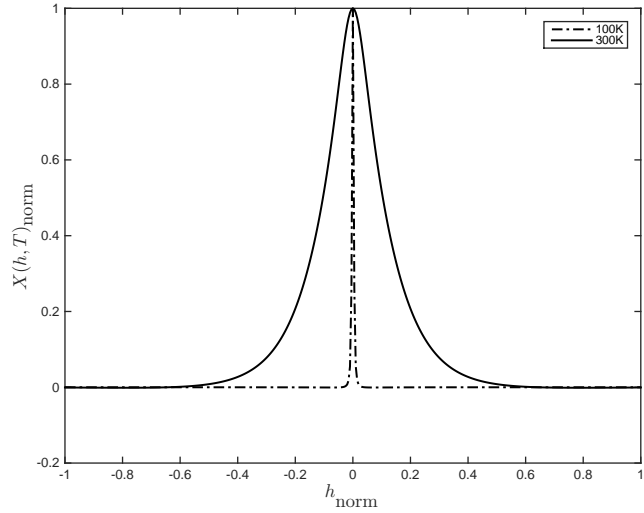}
 \caption{\footnotesize The susceptibility from the empirical PF. $\epsilon=1.0\times 10^{-2}\mbox{eV}$; $h$ is from $-0.02$ to $0.02\mbox{\ eV}$. The susceptibility and the external magnetic field are normalized by their maximum, respectively.}
\label{fig5}
\end{figure}

 The heat capacity, magnetization, susceptibility above can be shown intuitively in Fig. \ref{fig3}-\ref{fig5} through the numerical calculation.  It can be seen that specific heat approaches infinity at a certain temperature. The magnetization is discontinuous at $h=0$ when $T<T_c$. On the contrary, the magnetization is continuous at $h=0$ when $T>T_c$. Similar properties can also be seen for the susceptibility in Fig. \ref{fig5}.  

\begin{figure}[htb]
 \centering
 \includegraphics[scale=0.5]{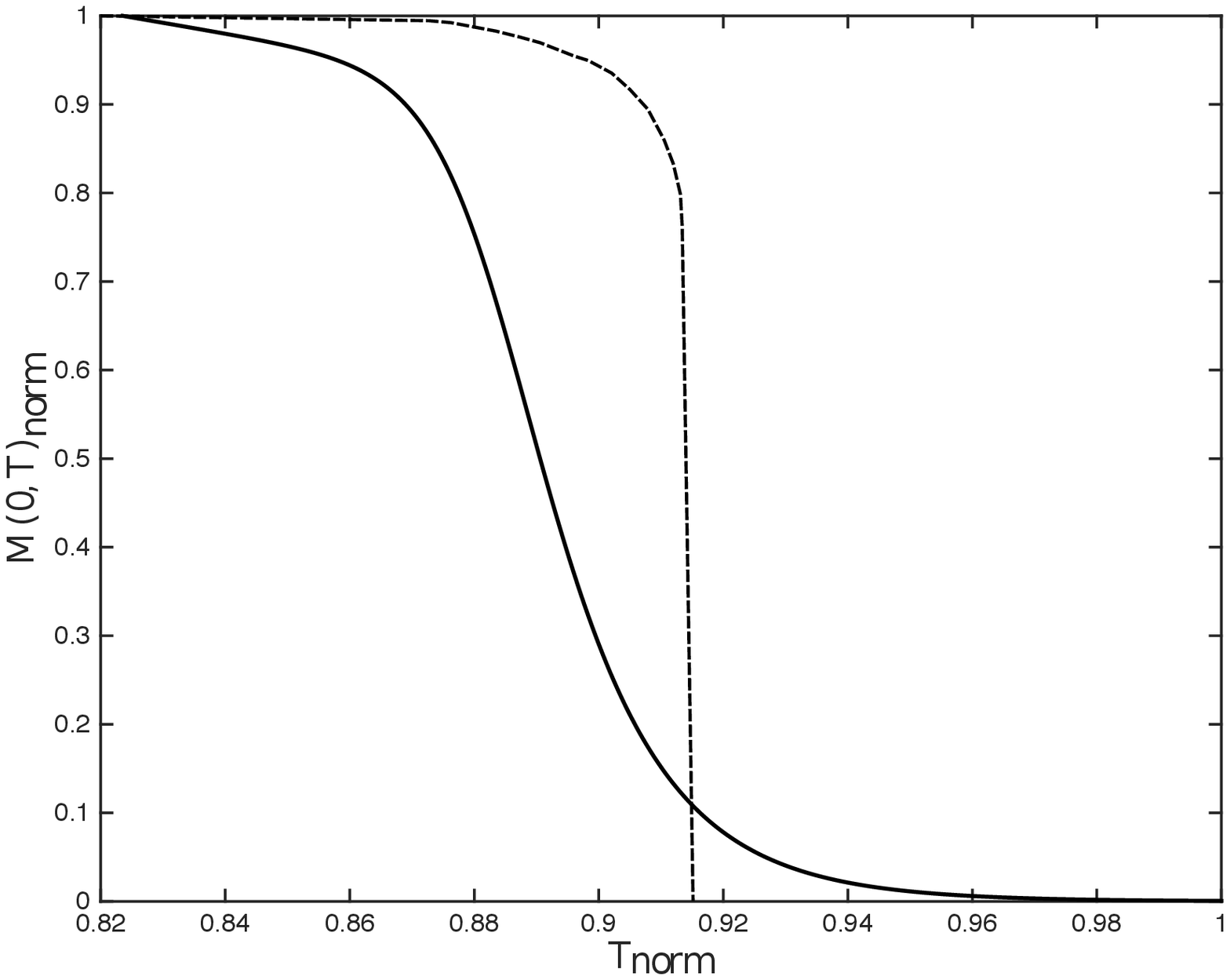}
 \caption{\footnotesize The spontaneous magnetization (SM) from the empirical PF and calculated with Eq. (\ref{magnetization}) (Solid line). $\epsilon=1.0\times 10^{-2}\mbox{eV}$; $h=1.0\times 10^{-30}\mbox{eV}$; $T$ is from 7.0 K to 8.5 K. The SM and temperature are normalized by their maximum, respectively. The dashed line is drawn qualitatively according to Yang (1952) for comparison. }
\label{fig6}
\end{figure}

Fig. \ref{fig6} shows the variation curve of the SM from the empirical PF with the temperature. It can be seen that this variation is different from that in Yang (1952). The curve reflects the general trend that the SM decreases with the increasing temperature. However, the normalized SM here taper towards 0 at the high temperature end, while at the low temperature end, it gradually approaches 1. This means that there is room for further improvement in our empirical PF. 

On the other hand,  the SM vs. temperature from the empirical PF is similar to those from the numerical simulation, for example, Steinberg et al.(2013), Secular (2015).

It should be pointed out that here Eq. (\ref{magnetization}) instead of Eq. (\ref{magnetization2}) is used for the sake of numerical calculation, in which we let $h=1.0\times 10^{-30}\mbox{eV}$ corresponding a zero magnetic field.

\subsection{Future work}
 
 As can be seen from the above, we can improve the empirical PF here.  A possible way is that we can modify the empirical PF based on the comparison with more exact results from the finite-size 2D Ising model. For example, such an exact PF can be from Wei (2018) with the improvement of computing power in the future. Especially, the SM from the improved empirical PF should be consistent with that of Yang (1952).

\section{Conclusions}

It is still a challenge to obtain a practicable and exact PF for the 2D Ising model in an external magnetic field, especially using the transfer-matrix method.  

Through the analysing of the PF for the 1D Ising model, and that for the 2D model with a zero external magnetic field, we infer an empirical PF for the 2D Ising model in an external magnetic field. The empirical PFs on two 2D Ising lattices are consistent well numerically with the results from Wei (2018). With this empirical PF, we found that: (1) the specific heat approaches infinity at a critical temperature logarithmically; (2) Both the magnetization and susceptibility show different behaviors when the temperature is above or below the critical temperature; (3) The 2D Ising model has the SM which decreases with the increasing temperature, implying that a phenomenon of the phase transition occurs at the critical temperature.

Our study shows that a reasonable empirical PF is also helpful for understanding the properties of the 2D Ising model with a non-zero magnetic field, although it has no rigorous derivation. With the improvement of the empirical PF, we can under the 2D Ising model better and better.

\vspace{5em}


\ \

\vspace{10em}

{\bf Appendix A.}

\ \ \ 

{\bf On "$\cosh^{-1}z=\frac{1}{\pi}\int_0^\pi \log(2z-2\cos\omega_1){\rm{d}}\omega_1$"}

\ \ \ \ 

Eq. (12) on page 535 in the book by Gradshteyn and Ryzhik (Gradshteyn, I. S., Ryzhik, I. M., 2014. Table of Integrals, Series, and Products (8th Edition), Academic Press (an imprint of Elsevier).) reads,

\begin{equation}
\int_0^{\pi}\log (1+a\cos x)\mbox{d}x=\pi\log \left( \frac{1+\sqrt{1-a^2}}{2} \right) \hspace{5em} a^2\le 1
\end{equation}

Clearly Eq. (\ref{wrq1}) holds,

\begin{equation}\label{wrq1}
\begin{array}{ll}
\int_0^{\pi}\log (1-a\cos x)\mbox{d}x & =\int_0^{\pi}\log [1+(-a)\cos x]\mbox{d}x \\
{}&=\pi\log \left( \frac{1+\sqrt{1-a^2}}{2} \right) \hspace{8em} a\ge 0 \mbox{\ and\ } a^2\le 1\\
{}&=\pi\log (1+\sqrt{1-a^2})-\pi\log 2
\end{array}
\end{equation}

Hence, 

\begin{equation}
\log (1+\sqrt{1-a^2})=\frac{1}{\pi}\int_0^{\pi}\log (1-a\cos x)\mbox{d}x+\log 2
\end{equation}

That is,

\begin{equation}
\log (1+\sqrt{1-a^2})=\frac{1}{\pi}\int_0^{\pi}\log (1-a\cos \omega_1)\mbox{d}\omega_1+\log 2
\end{equation}

Therefore,

\begin{equation}
\begin{array}{ll}
{\cosh ^{ - 1}}z &= \log (z + \sqrt {{z^2} - 1} ) \hspace{2em} {(z\ge 1)}\\
{}&= \log z +\log (1 + \sqrt {1 - {z^{ - 2}}} )\\
{}&=\log z +\frac{1}{\pi}\int_0^{\pi}\log (1-\frac{1}{z}\cos \omega_1)\mbox{d}\omega_1+\log 2\\
{}&=\frac{1}{\pi}\int_0^\pi \log(2z-2\cos\omega_1){\rm{d}}\omega_1
\end{array}
\end{equation}


\begin{thebibliography}{99}
\bibitem{}
Dixon, J. M., Tuszynski, J. A., Nip, M. L. A., 2001. Exact eigenvalues of the Ising Hamiltonian in one-, two- and three-dimensions in the absence of a magnetic field, Physica A, 289: 137-156.
\bibitem{}
Houtappel, R.M.F., 1950. Order-disorder in hexagonal lattices. Physics XVI, 5:425-455.
\bibitem{}
Huang K., 1987. Statistical Mechanics (2nd Edition),  John Wiley \& Sons, Inc..
\bibitem{}
Joyce, G.S., 2003. Singular behaviour of the lattice Green function for the d-dimensional hypercubic lattice, J. Phys. A: Math. Gen. 36: 911–921.
\bibitem{}
Kramers, H.  A.,  Wannier, G. H.,  1941. Statistics of the Two-Dimensional Ferromagnet. Part I.  Phys. Rev.,  60: 252-262.
\bibitem{}
Mccoy, B.M., Wu, T.T., The two-dimensional Ising model, Harvard University Press, Cambridge, Massachusetts, 1973.
\bibitem{}
Martin, P.  P.,  Potts Models and Related Problems in Statistical Mechanics, volume 5. World Scientific, 1991.
\bibitem{}
Onsager, L., 1944. Crystal Statistics. I. A Two-Dimensional Model with an Order-Disorder Transition. Phys. Rev.,  65(3-4): 117-149.
\bibitem{}
Secular, P., 2015. Monte-Carlo simulation of small 2D Ising lattice with Metropolis dynamics, http://secular.me.uk/physics/ising-model.pdf
\bibitem{}
Steinberg, A.P., Kosowsky, M., Fraden, S., 2013. Simulations: The Ising Model, http://216.92.172.113/courses/phys39/simulations/AsherIsingModelReport.pdf
\bibitem{}
Wei, R.Q., 2018. An exact solution to the partition function of the finite-size Ising Model, arXiv: General Physics: 1805.01366.
\bibitem{}
Yang, C.N., 1952. The spontaneous magnetization of a two-dimensional Ising Model, Physical Review, 85(5): 808-816.
\end{thebibliography}
\end{document}